\documentclass[aps,pre,twocolumn,floatfix,showpacs,showkeys]{revtex4}
\usepackage{here}
\usepackage{graphicx}
\usepackage{bm}% bold math

\newcommand\pictc[5]{\begin{figure}
                       \centerline{
                       \includegraphics[width=#1\columnwidth]{#3}}
                   \protect\caption{\protect\label{fig:#4} #5}
                    \end{figure}            }
\newcommand\pict[4][1.]{\pictc{#1}{!tb}{#2}{#3}{#4}}

\newcounter{Fig}

\begin{document}

\title{Sub-wavelength imaging with a left-handed material flat lens}

\author{Michael W. Feise}
\affiliation{Nonlinear Physics Group and Centre for Ultra-high
bandwidth Devices for Optical Systems (CUDOS), Research School of
Physical Sciences and Engineering, Australian National University,
Canberra, ACT 0200, Australia}

\author{Yuri S. Kivshar}
\affiliation{Nonlinear Physics Group and Centre for Ultra-high
bandwidth Devices for Optical Systems (CUDOS), Research School of
Physical Sciences and Engineering, Australian National University,
Canberra, ACT 0200, Australia}

\date{\today}

\begin{abstract}
  We study numerically, by means of the pseudospectral time-domain
  method, the unique features of imaging by a flat lens made of a
  left-handed metamaterial that possesses the property of negative
  refraction.
  We confirm the earlier finding that a
  left-handed flat lens can provide near-perfect imaging of a point
  source and a pair of point sources with clear evidence of the
  sub-wavelength resolution. We illustrate the limitation of the
  resolution in the time-integrated image due to the presence of
  surface waves.
\end{abstract}



\maketitle

The unique properties of left-handed (LH)
materials~\cite{Veselago:1968-509:PUS}, i.e. materials with
simultaneously negative real parts of the dielectric permittivity
$\epsilon_r$ and the magnetic permeability $\mu_r$ (which can also be
described by a negative index of refraction), allow focusing of
electromagnetic waves by a flat slab of the material; this effect is
in sharp contrast to conventional optical lenses with a positive
refractive index that need to have curved surfaces to form an image.
Recently, Pendry~\cite{Pendry:2000-3966:PRL} argued that a slab of a
lossless LH material with $\epsilon_r = \mu_r =-1$ should behave as a
perfect lens enabling one to obtain an ideal image of a point source
through the reconstitution of the evanescent wave components. While
recent experiments confirmed the main features of negative
refraction~\cite{Shelby:2001-77:SCI,Parazzoli:2003-107401:PRL}, the
question of near-perfect imaging of a flat lens and near-field
focusing has remained highly controversial~\cite{Venema:2002-119:NAT},
and is severely constrained because of large dissipation and
anisotropy in the metamaterial.  Nevertheless, several studies showed
that nearly-perfect imaging should be expected even under realistic
conditions when both dispersion and losses of the left-handed material
are taken into
account~\cite{Fang:2003-161:APL,Cummer:2003-1503:APL,Smith:2003-1506:APL,Lu:2003-282:MOTL,Nieto-Vesperinas:2004-491:JOSA}.

In this Letter, we re-visit the problem of nearly perfect imaging by a
flat lens made of a left-handed metamaterial and study numerically, by
use of the pseudospectral time-domain method, imaging by a flat lens
made of this material. In order to study the amplification of the
evanescent waves, we compare the wave-vector spectra of the field at
the image plane in the case with and without the LH slab.  We confirm
the earlier finding that a left-handed flat lens can provide
near-perfect imaging of a point source~\cite{Cummer:2003-1503:APL} and
a pair of point sources with clear evidence of sub-wavelength
resolution. In addition we consider the time-resolved and
time-integrated Poynting vector at the source frequency which behaves
equivalently to the time-integrated intensity that may be relevant in
lithography applications.

\pict{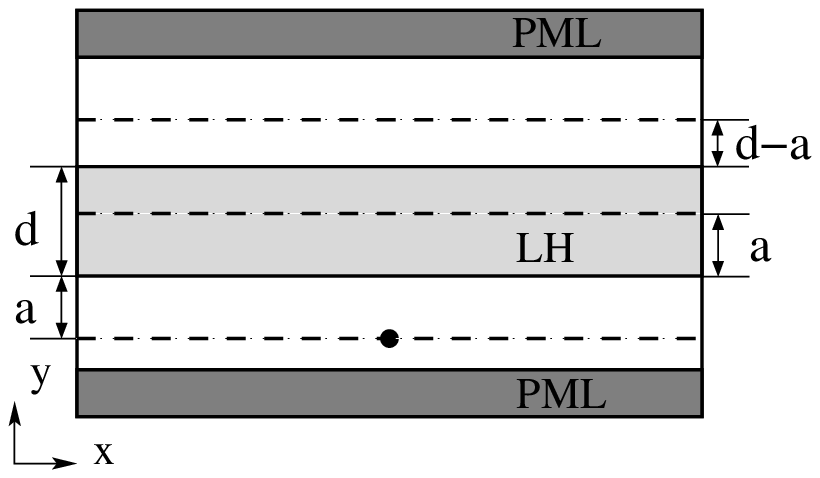}{schematic-structure}{Schematic view of the model
  system. A current source is placed in front of a LH slab. In $y$
  direction the system models an open domain through the use of
  reflectionless absorbing boundaries (PML), while it is periodic in
  the $x$ direction. The solid dot indicates the source location while
  the dashed lines denote the observation planes, i.e.\ the source and
  image planes.}

We model an electrical line current source in front of a slab of LH
material embedded in free space, as shown in
Fig.~\ref{fig:schematic-structure}. The system is translationally
invariant in the $z$ direction and is treated as two-dimensional.  The
simulations are performed in TE polarization ($H_x, H_y, E_z \neq 0$).
In the $y$ direction the system is considered open, as achieved
through reflectionless absorbing boundaries, while the $x$ direction
is taken as periodic. Because of this the simulation essentially uses
an infinite slab and an array of sources.

It has been shown that a material with left-handed character can only
be achieved through dispersion \cite{Veselago:1968-509:PUS}.
Therefore, the LH material is assumed to have lossy Drude
characteristics in both electric permittivity and magnetic
permeability \cite{Pendry:2000-3966:PRL}, given by
\begin{eqnarray}
  \label{eq:1}
  \varepsilon_r(\omega) &=&
  1-\frac{\omega_{pe}^2}{\omega(\omega +i\gamma_e)},\\
  \label{eq:2}
  \mu_r(\omega) &=& 1-\frac{\omega_{pm}^2}{\omega(\omega+i\gamma_m)}.
\end{eqnarray}
Here $\omega_{pe}$, $\omega_{pm}$ are the plasma frequencies and
$\gamma_e$, $\gamma_m$ are the collision frequencies of the electric
and magnetic properties, respectively. To simplify and to impedance
match the slab to free space, we take
$\omega_{pe}=\omega_{pm}=\omega_p$ and $\gamma_e=\gamma_m=\gamma$. The
material parameters are chosen to give a refractive index with real
part $\mathrm{Re}(n)=-1$ at frequency $f_0=15$~GHz ($\omega_0=2\pi
f_0$). For this we use $\omega_p=2\pi\sqrt{2}f_0$. The collision
frequency is $\gamma=2\pi\times 4.5$~MHz, which results in
$\varepsilon_r(\omega_0)=\mu_r(\omega_0)=-1+0.0006i$.

We directly simulate the field propagation based on Maxwell's
equations, using the pseudospectral time-domain
method~\cite{Liu:1997-158:MOTL}. In this method all field components
are sampled at the same spatial location which avoids ambiguities of
the material interfaces~\cite{Feise:2004-2955:ITAP} inherent in the
usual finite-difference time-domain method
\cite{Yee:1966-302:ITAP,Taflove:2000:ComputationalElecetromagnetics},
and the problems caused by transition layers at the interfaces
\cite{Feise:2002-35113:PRB}.  The domain walls in the $y$ direction
are covered with a uniaxial perfectly-matched layer (PML)
\cite{Berenger:1994-185:JCP,Sacks:1995-1460:ITAP,Kuzuoglu:1996-447:IMGL}
boundary to simulate an open domain. In the $x$ direction the system
is periodic.  The material constitutive relations are implemented
using the auxiliary differential equation method
\cite{Joseph:1991-1412:OL}.

The current source is turned on slowly using a temporal dependence
$(1-e^{-t/\tau})\sin(\omega_0 t)$, with a turn-on parameter
$\tau\approx22/f_0$, to reduce its band width and allow the system to
reach steady-state faster.

The simulation uses a time step of $\Delta_t = 29.2873$~ns and a
spatial step size of $\Delta_x = \Delta_y = \lambda_0/100 =
0.199862$~m; $\lambda_0$ is the free-space propagating wavelength at
frequency $f_0$.  The simulation size is $1024\times 128$ and is
iterated for 600000 time steps, i.e.\ 2635 periods, to ensure
steady-state.

It has been shown that a slab of LH material with parallel sides
should focus propagating \cite{Veselago:1968-509:PUS} and evanescent
\cite{Pendry:2000-3966:PRL} waves. Thus, if the slab thickness $d$ is
greater than the distance $a$ between the object (source) and the
front face of the slab, then one expects an image to form inside the
slab, a distance $a$ from the front face, as well as behind the slab,
a distance $d-a$ from the back face, as indicated in
Fig.~\ref{fig:schematic-structure}.

\pict{fig02.eps}{field-single}{Snapshots of
  the electric field in the source plane (solid), the image plane
  inside the slab (dotted) and beyond the slab (dashed) with a single
  source.}

\pict{fig03.eps}{field-wavevector-single}{Transverse wave
  vector dependence of the fields in the image planes relative to the
  field in the source plane in the case of a single source; with LH
  slab: image location inside slab (dotted), image location beyond
  slab (dashed); free-space: image location inside slab (dash-dotted),
  image location beyond slab (solid). }

First, we place a single source a distance $a=0.2 \lambda_0$ in front
of a slab with thickness $\lambda_0/3$. Snapshots of the electric field
in the source plane and the two image planes are shown in
Fig.~\ref{fig:field-single}. The snapshots were taken when the field
was at its maximum in the source plane, which was also the time of
maximum field in the image planes. The field of the source plane is
well reproduced in the image planes. The feature size of the central
peak in the image planes is well below the free-space limit of
$\lambda_0/2$. To illustrate that point we performed a spatial Fourier
transform of this data and show the transverse spatial frequency
dependence of the image-plane fields relative to the source-plane
field in Fig.~\ref{fig:field-wavevector-single}. For comparison the
figure also shows the spectrum in the same planes when the LH slab is
replaced by air. In the case without the LH slab, waves with spatial
frequency greater than $1/\lambda_0$, corresponding to evanescent
waves, are almost entirely absent in the image planes. In contrast,
when the LH slab is present, waves with spatial frequency up to
$3.5/\lambda_0$ are transmitted to the image locations.  This agrees
with the smaller than $\lambda_0/2$ features present in
Fig.~\ref{fig:field-single}.  The wave vector cutoff value
($k_c=2\pi\lambda_c$) seen in Fig.~\ref{fig:field-wavevector-single}
is comparable to the value expected according to the parameters used
in the system \cite{Smith:2003-1506:APL}. At $2.8/\lambda_0$ a peak
occurs in the spectrum, similar to the results reported in
\cite{Rao:2003-67601:PRE}.

\pict{fig04.eps}{Poynting-double}{Snapshot of the
  $y$ component of the Poynting vector in the image plane beyond the
  slab for a slab thickness of $\lambda_0/3$ (solid) and $\lambda_0/2$
  (dashed). }

We also study the imaging of a pair of sources in order to
characterize the possibility of sub-wavelength resolution. The single
source is replaced by a pair, separated by $0.34\lambda_0$. The two
sources have the same time dependence, i.e.,\ are in-phase. In
Fig.~\ref{fig:Poynting-double} snapshots of the $y$ component of the
Poynting vector (${S}_y$) in the image plane beyond the slab
are shown for two different slab thicknesses while the distance
between the sources and the slab is keept constant. The snapshots are
again taken at the same time and show the instance with highest image
peaks. One finds that features with sizes less than $\lambda_0/2$ are
reproduced in the images but the quality of the sub-wavelength
resolution diminishes with increased slab thickness
\cite{Smith:2003-1506:APL}.

With slab thickness $\lambda_0/3$, ${S}_y$ in the region
between the image peaks has opposite sign compared to that of the
image peaks.  This is a signature of the influence of the surface
waves on the image.

\pict{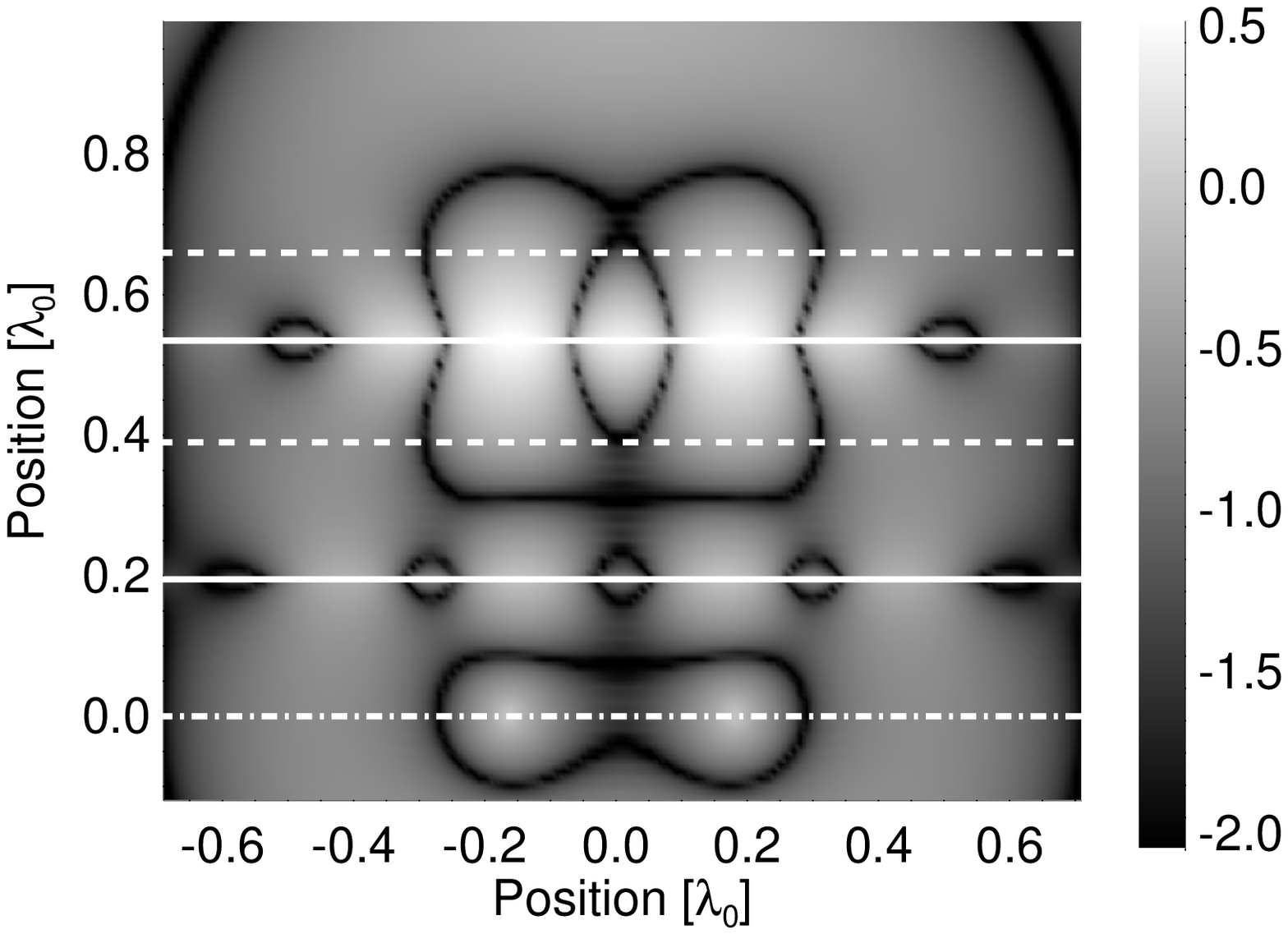}{2-d-snapshot-double}{Absolute magnitude of the
  electric field on a logarithmic scale at time $t=2635/f_0$ with two
  sources in front of the LH slab of thickness $\lambda_0/3$.  The LHM
  slab is denoted by solid lines, the image planes by dashed lines,
  and the source plane by a dash-dotted line.}

Figure~\ref{fig:2-d-snapshot-double} shows a two-dimensional
false-color plot of the absolute magnitude of the electric field with
two sources on a logarithmic scale. The geometry and orientation is as
indicated in Fig.~\ref{fig:schematic-structure}.  One clearly sees the
two sources and the strong fields they excite at the surfaces.  The
surface waves rapidly diminish within $\lambda_0/2$ on either side of
the source--image axis.  One notices that in the image planes the
source images are resolved in the transverse, i.e.\ $x$, direction but
not in the $y$ direction. Again, we see the additional peak between
the image peaks and extrema of the surface waves to both sides.

To understand the behavior of the surfave waves we studied the
transfer function of the system, which is easily found by matching the
boundary conditions of the fields at the slab surfaces. With the
parameters of the system given above, we found that the appearance of
the surface waves is dominated by the high wave vector components
contained in the source distribution. These wave vector components lie
beyond the maximum wave vector value $k_c$ that the slab can image
correctly. A source distribution with less contribution beyond $k_c$
suffers less distortion in the image from surface waves. The slab
thickness and the material parameters determine $k_c$ which in turn
influences the occurrance of the surface waves.

Unlike the cavity effect observed for a slab of finite width
\cite{Chen:2004-107404:PRL}, we found that the distance to the
neighboring sources (present due to the periodic boundary conditions
in $x$) has no noticable influence on the image, as long as this
distance is at least a few $\lambda_c$.

\pict{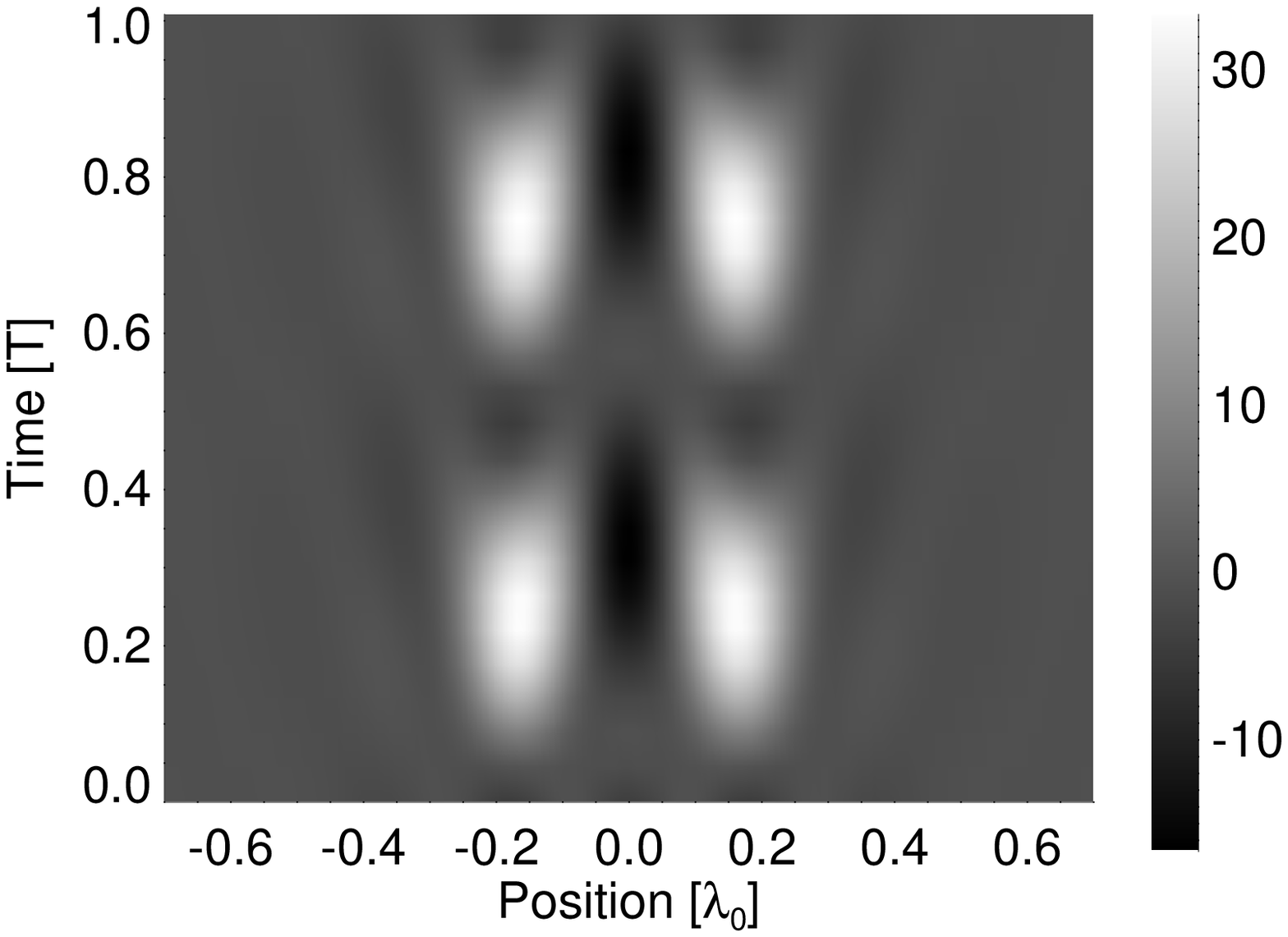}{Poynting-time-lambda_3}{Time
  dependence of the amplitude of the $y$ component of the Poyting
  vector (${S}_y$) in the image plane beyond a slab of
  thickness $\lambda_0/3$ on a linear scale.}

To illustrate the temporal behavior of the image, in
Fig.~\ref{fig:Poynting-time-lambda_3} we plot ${S}_y$ in the
image plane beyond the slab as a function of time for a slab of
thickness $\lambda_0/3$.  The image peaks are visible as areas of
large positive value. Between the image peaks, with a time delay of
about $T/8$, a negative peak appears. The curve in
Fig.~\ref{fig:Poynting-double} corresponds to a slice at time $0.22 T$
in Fig.~\ref{fig:Poynting-time-lambda_3}.  Upon careful inspection one
can see the peaks of the surface waves moving away from the image
peaks and then disappear. When the slab thickness is increased to
$\lambda_0/2$ (not shown), the region of negative ${S}_y$ has
essentially disappeared and become a region of relatively large
positive $S_y$. Rather than being clearly distinct, the two image
peaks begin to merge, as can be seen in Fig.~\ref{fig:Poynting-double}.
The surface waves present have lower $k_c$, i.e., larger $\lambda_c$,
and greater amplitude. They also move faster away from the image
peaks.

From Fig.~\ref{fig:Poynting-time-lambda_3} it is evident that the
time-integrated Poynting vector has an additional significant extremum
between the two image peaks, which will lead to a distortion of the
image.  In general, this peak due to surface waves is not
distinguishable from a genuinely imaged peak in the source plane.
This poses an important limitation on the usable parameter range of
the left-handed material flat lens.

In conclusion, we have studied the unique features of imaging by a
flat lens made of a left-handed metamaterial that possesses the
property of negative refraction.  By employing the pseudospectral
time-domain method and comparing with the wave propagation in air, we
have confirmed previous results of near-perfect imaging of a point
source and a pair of point sources with clear evidence of
sub-wavelength resolution. We found that the time-integrated image has
an additional limiting factor on the resolution of the images of
multiple close objects due to substantial constructive interference in
the image plane.  We believe that a potential advantage of this kind
of imaging is its scalability to sub-micrometer scales to make
possible imaging and nano-photolithography with spatial resolution in
the tens of nanometers.

This work has been partially supported by the Australian Research
Council and the Australian National University. The authors thank I.\ 
V.\ Shadrivov for helpful collaboration, N.\ A.\ Zharova and
A.\ A.\ Zharov for useful discussion of left-handed materials,
as well as S.\ A.\ Cummer, J.\ B.\ Schneider and R.\ W.\ Ziolkowski
for useful discussion of numerical methods.

\end{document}